\documentclass{svproc}
\usepackage{url}
\usepackage{cite}
\usepackage{amsmath,amssymb,amsfonts}
\usepackage{algorithmic}
\usepackage{graphicx}
\usepackage{textcomp}
\usepackage{multirow}
\usepackage{xcolor}
\usepackage{adjustbox}
\usepackage[english]{babel}
\usepackage[T1]{fontenc}
\usepackage[utf8]{inputenc}

\begin{document}
\mainmatter              
\title{Semantic Sensitive TF-IDF to Determine Word Relevance in Documents}
\titlerunning{Semantic Sensitive TF-IDF}  
%
\author{Amir Jalilifard\inst{2} \and Vinicius Fernandes Caridá\inst{1} \and Alex Fernandes Mansano \inst{1} \and Rogers S. Cristo \inst{1} \and Felipe Penhorate Carvalho da Fonseca \inst{1}}
\authorrunning{Amir Jalilifard et al.} 
%
\tocauthor{Amir Jalilifard,  Vinicius F. Caridá, Alex F. Mansano, and Felipe P.C. Fonseca}
\institute{Data Science Team - Digital Customer Service, Itaú Unibanco, São Paulo, Brazil\\
\and
Federal University of Minas Gerais, Brazil\\
\email{jalilifard@ufmg.br and vinicius.carida; alex.mansano; rogers.cristo; felipe.fonseca(@itau-unibanco.com.br)}\\ 
}

\maketitle              

\begin{abstract}
Keyword extraction has received an increasing attention as an important research topic which can lead to have advancements in diverse applications such as document context categorization, text indexing and document classification. In this paper we propose STF-IDF, a novel semantic method based on TF-IDF, for scoring word importance of informal documents in a corpus. A set of nearly four million documents from health-care social media was collected and was trained in order to draw semantic model and to find the word embeddings. Then, the features of semantic space were utilized to rearrange the original TF-IDF scores through an iterative solution so as to improve the moderate performance of this algorithm on informal texts. After testing the proposed method with 160 randomly chosen documents, our method managed to decrease the TF-IDF mean error rate by a factor of 50\% and reaching the mean error of 13.7\%, as opposed to 27.2\% of the original TF-IDF. \dots
\end{abstract}
\section{Introduction}
In the information era when huge number of digital documents are gathered in a daily basis, going through documents and extracting the most relevant information, understanding the general concept and finding the other related documents is more necessary than ever. Keywords are several relevant words that provide a rich semantic information about a text for many natural language processing applications. Thereby, many researches have been carried out in order to extract the most relevant words from a text. 

Some made use of the already-known supervised classification methods such as Support Vector Machine (SVM) and Naive Bayes \cite{zhang2006keyword} \cite{uzun2005keyword}. Although these supervised approaches methods provided good results, the need for training data, which often needs involving human resources, still remain a problem. Moreover, the word relevance score provided by a method like SVM may or may not be directly proportional to the importance of terms in a particular document. Therefore, an unsupervised method which provide local weights considering a class of documents is desirable.

Term frequency–inverse document frequency (TF-IDF) is one of the most commonly used term weighting schemes in information retrieval systems. Despite its popularity, TF-IDF has often been considered an empirical method, specifically from a probabilistic point of view, with many
possible variations\cite{Aizawa2003}. TF-IDF is a numerical statistics that, by scoring the words in a text, indicates how important a word is in a document considering the corpus that document belongs to. This method was studied in several researches for keyword and word relevance extraction. 

\begin{figure}[!ht]
  \centering
  \includegraphics[width=8cm]{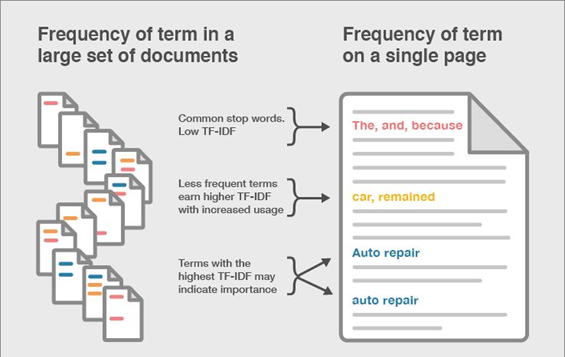}
  \caption{A simple example of TF-IDF. Adapted from: https://moz.com/blog/the-technical-seo-renaissance}
  \label{fig:extfidf}
\end{figure}

In this paper, we propose STF-IDF, a novel semantic sensitive method based on the conventional TF-IDF. The key idea is readjusting the conventional TF-IDF scores based on the semantic representation of most relevant words. Thereby, we assume that if a set of terms is considered important by TF-IDF, all the semantically similar words related to this set should be considered more important than those ones with less semantic relevance to the context. 
The next section explains the theoretical basis of TF-IDF and discuss about the related works. In the section three is explained the proposed method. The results and discussion are presented in the last section.

\section{Related Works}\label{sec:Related Works}

The Vector Space Model, generally attributed to Salton et al. \cite{Salton1975} and stemming from the Information Retrieval community, is arguably the most successful and influential model to encode words and documents as vectors \cite{almeida2019word}. Salton et al. \cite{Salton1975} suggest an encoding procedure whereby each document in a collection is represented by a t-dimensional vector, each element representing a distinct term contained in that document. These elements may be binary or real numbers, optionally normalized using a weighting scheme such as TF-IDF, to account for the difference in information provided by each term.

\subsection{An Overview of TF-IDF}\label{sec:An Overview of TF-IDF}

Inverse document frequency (IDF) \cite{Papineni2001} is one of the most important and widely used concepts in information retrieval. It is used in combination with the term frequency (TF). The result is a very effective term weighting scheme that has been applied for information retrieval systems \cite{Singh2014}. Essentially, TF-IDF works by determining the relative frequency of words in a specific document compared to the inverse proportion of that word over the entire document corpus. Intuitively, this calculation determines how relevant a given word is in a particular document. Words that are common in a single or a small group of documents tend to have higher TF-IDF numbers than common words such as articles and prepositions\cite{Ramos2003UsingTT}. The formal procedure for implementing TF-IDF has some minor differences over all its applications, but the overall approach works as follows: given a document collection D, a word w, and an individual document \begin{math} d \epsilon D\end{math}, calculate:

\begin{equation}
    w_{d} = f_{w,d} * log (|D|/f_{w,D})
\end{equation}

Where \begin{math} f_{w,d} \end{math} equals the number of times w appears in d, |D| is the size of the corpus, and \begin{math} f_{w,D} \end{math} equals the number of documents in which w appears in D \cite{Salton1988} \cite{Berger2000}. There are a few different situation that can occur here for each word, depending on the values of \begin{math} f_{w,d} \end{math} |D|, and \begin{math} f_{w,D} \end{math}, the most prominent of which we'll examine. 
Assume that \begin{math}|D| ~ f_{w,d} \end{math} i.e. the size of the corpus is approximately equal to the frequency of w over D. If \begin{math} 1 < log (|D|/ f_{w,D}) \end{math}  for some very small constant c, then \begin{math}  w_{d} \end{math} will be smaller than \begin{math} f_{w,d} \end{math} but still positive. This implies that w is relatively common over the entire corpus but still holds some importance throughout D. For example, this could be the case if TF-IDF would examine the word "Jesus" over the New Testament. More relevant to us, this result would be expected of the word "United" in the corpus of United Nations documents. This is also the case for extremely common words such as articles, pronouns, and prepositions, which by themselves hold no relevant meaning in a query (unless the user explicitly wants documents containing such common words). Such common words thus receive a very low TF-IDF score, rendering them essentially negligible in the search \cite{Ramos2003UsingTT}.
Finally, suppose \begin{math}fw, d \end{math} is large and \begin{math} fw, D \end{math} is small. Then \begin{math} log (|D|/ fw, D) \end{math} will be rather large, and so \begin{math} w_{d} \end{math} will likewise be large. This is the case we're most interested in, since words with high \begin{math} w_{d} \end{math} imply that w is an important word in d but not common in D. This w term is said to have a large discriminatory power. Therefore, when a query contains this w, returning a document d where \begin{math} w_{d} \end{math} is large will very likely satisfy the user.

The code for TF-IDF is elegant in its simplicity. Given a query q composed of a set of words \begin{math} w_{i} \end{math}, we calculate \begin{math} w_{i,d} \end{math} for each \begin{math} w_{i} \end{math} for every document \begin{math} d \epsilon D\end{math}. In the simplest way, this can be done by running through the document collection and keeping a running sum of \begin{math} f_{w,d} \end{math} and \begin{math} f_{w,D} \end{math}.
Once done, we can easily calculate \begin{math} w_{i d} \end{math}  according to the mathematical framework presented before. Once all \begin{math} w_{i d} \end{math}'s are found, we return a set D* containing documents d such that we maximize the following equation

\begin{equation}
    \sum_{i} w_{i,d}
\end{equation}

Either the user or the system can arbitrarily determine the size of D* prior to initiating the query. Also, documents are returned in a decreasing order according to equation (2). This is the traditional method of implementing TF-IDF \cite{Ramos2003UsingTT}.

\subsection{Related Works}\label{sec:Related Works}

TF-IDF, in general, is one of the most used techniques to quantify a word in documents. However, there are variations in some methods. Singh and Dwivedi \cite{Singh2014} compared four methods of IDF found in literature using a set of TREC based queries. After computing the value of the IDF using keyword based search, they concluded that the IDF value computed by \cite{YJung2000} gives better TF-IDF weight of terms compared to other methods. 

Ramos and his colleagues \cite{ramos2003using} examined the result of applying TF-IDF in determining the word relevance in document queries and concluded that this simple method efficiently classifies relevant words. Li et al. \cite{li2007keyword} applied TF-IDF for keyword extraction in Chinese texts based on analyzing linguistic characteristics of documents and providing several strategies including uni-, bi- and tri-gram extraction, new word finding and refinement. Chung et al. \cite{wu2008interpreting} proposed a probabilistic model based on TF-IDF which makes local relevance decisions for each location in a document and combines these local relevant decisions into a document wide relevance decision. Lee and colleagues \cite{lee2008news} presented several variants of conventional TF-IDF for a more effective keyword extraction and topic summarization. They used cross-domain comparison for removing meaningless or irrelevant words.

Despite the advances in word embeddings (i.e. word vector representations), capturing sentence meaning is an open question due to complexities of semantic interactions among words \cite{Ignacio2019}. Even papers that use embeddings, some times, use TF-IDF to help to fitted The weights of the series. Another challenge in document classification is insufficient label information and sparse, unstructured format. With the intention of increasing the variety of resource sets for classification, the authors \cite{Donghwa2019} used three methods of document representation, including TF-IDF and proposed multi-co-training (MCT). In order to help the categorization of documents into hierarchical structures showing the relationship between variables \cite{Iwendi2019} proposes a novel TF-IDF algorithm with the temporal Louvain approach. 

Although the aforementioned methods improved the performance of conventional TF-IDF by providing probabilistic solutions or the use of multi-strategies, they are highly dependent on the original TF-IDF idea which is giving more weight to the words with high local and less global probability. This consideration specially fails when it comes to finding word relevance in informal documents, which are important sources of information in the era of social networks \cite{morris2010comparison}. As an example, if a text contains informal words related to a specific ethnic communities, or words that have been used in a specific period of time, but not in the whole corpus, due to some changes in cultural expressions, both conventional TF-IDF and the related methods that attempt to improve its performance fail to find relevant words with high accuracy. Another example is informal conversations regarding formal topics like medical communities that provide rich information for users. Censuring the general semantic context, TF-IDF fails to detect the context-sensitive content which plays an important role in informal texts \cite{wollmer2010bidirectional}.

\section{Materials and method}

In this section we explain the materials and the mathematical definition of our proposed method. We start with the data acquisition and then we explain how our algorithm tries to improve the conventional TF-IDF through a finite numbers of iterations.

Data of nearly four million pages of online medical communities was gathered and after being pre-processed (i.e removing stopwords, punctuation, etc.), they were fed to word2vec \cite{mikolov2013efficient} in order to learn the semantic space and words' distribution. Having the semantic distribution of terms of the corpus, our algorithm generates the word relevance score as following:


\begin{equation}
    S_{wj}^{(k)} = S_{wj}^{(k-1)} \times \frac{1}{1 + \left \| e(w_j) \right \|\left \| \overline{e(w)}) \right \| cos(\Theta) }
\end{equation}
where \begin{math} S_{wj}^{(k)} \end{math} is the vector of word scores in \begin{math} Kth \end{math} iteration and the initial scores are calculated using the conventional TF-IDF:
\begin{equation}
    S_{wj}^{(0)} = P(W_j)_d * Log(P(W_j)_c) = TFIDF_{Wj}
\end{equation}

and \begin{math} \overline{e(w)} \end{math} is the weighted expected value of the first \begin{math} \left \lfloor \sqrt{n} \right \rfloor \end{math} most relevant words in the previous iteration and is calculated as follow:

\begin{equation}
      \overline{e(w)} = \frac{1}{{\left \lfloor \sqrt{n} \right \rfloor} }\sum_{i=1}^{\left \lfloor \sqrt{n} \right \rfloor}(\frac{1}{1 - \frac{S_{wj}^{(k-1)}}{\sum_{j=1}^{n}S_{wj}^{(k-1)}}} \times  e(w_{j}))
\end{equation}

The algorithm is initiated with conventional TF-IDF scores. In each iteration, first, the mean embedding of \begin{math} \left \lfloor \sqrt{n} \right \rfloor \end{math} most relevant words from previous iteration are selected and the weighted mean embedding of them is calculated. The idea is that words with higher scores represent the context more than those which are less relevant. In order to calculate the weighted mean embedding, for each word its score in the previous iteration is divided by all the scores in order to get a number in the range [0,1]. This weight then is converted to a number greater than 1 and is considered the weight by which each word pushes the mean embedding toward itself. Afterwards, the previous scores are recalculated by being multiplied on the cosine distance of the word and the mean embedding. The idea is to repeatedly replace the words that have poor representation of general text context with those that are more related to the document context. The new scores are then passed to the next iteration and the scores are rearranged so that the words with better representation are moved toward the top of ranking.

By considering the embedding of the first \begin{math} \left \lfloor \sqrt{n} \right \rfloor \end{math} most relevant words as a multivariate unknown probability distribution, the less-relevant words are those instances which increase the variance of the distribution. Therefore, the goal is to decrease the variance by constantly replacing the outliers with those words that are semantically more related to the most important words and move the expected value of embeddings toward the value that perfectly matches the context. First, we prove that in each iteration the variance of the distribution from the mean in the previous iteration is decreased. Then, it is shown that in each iteration the mean value of set moves toward the mean of ideal distribution until it converges.

We define the mean embedding of the observed data in each iteration and the unknown distribution that best fits the context with \begin{math} \mu^{(i)} \end{math}, and \begin{math} \mu \end{math}, respectively, as following:
\begin{equation}
\mu^{(i)} = \frac{X^{(i)}_{1}+X^{(i)}_{2}+X^{(i)}_{3}+ ... + X^{(i)}_{m-1} + X^{(i)}_{m}}{m}
\end{equation}
where \begin{math} m = \left \lfloor \sqrt{n} \right \rfloor \end{math} and \begin{math} X_{m}^{(i)} \end{math} represents the embedding of \begin{math} m \end{math}th word of the set in the \begin{math} i \end{math}th iteration and \begin{math} \mu^{(0)} \end{math} as the mean of the initial distribution generated by conventional TF-IDF, and \begin{math} X_{m}^{(i)} \neq X_{m}^{(i-1)} \end{math} if a word is substituted.

In each iteration either the set maintains the current members or some of them are replaced with words that have a less cosine distance. Multiplying the word score from previous iteration \begin{math} S_{wj}^{(k-1)} \end{math} to the inverse of cosine distance guarantees that a word with higher cosine distance from the expected value of embeddings is decreased more than a word with less distance. A factor of 1 is added to the equation in order to eliminate the reverse effect of distances lesser than 1. 

In the simplest case, we assume that in each iteration one word is replaced with another. As it was explained, the equation (3) guarantees that the new word has less cosine distance to the expected value of embeddings than the replaced word. As a result: 

\begin{equation}
    \frac{1}{m}\sum_{k=1}^m (X^{(i)}_{k} - \mu^{(i-1)})^2 < \frac{1}{m}\sum_{k=1}^m (X^{(i-1)}_{k} - \mu^{(i-1)})^2 \Rightarrow Var(X^{(i)}) < Var(X^{(i-1)})
\end{equation}

\begin{equation} 
    \label{eq:6}
    \Rightarrow
    \lim_{i \rightarrow \infty} \frac{Var(X^{(i-1)})}{Var(X^{(i)})} = 1  
\end{equation}

Let's assume that the real context mean is bigger than the initially estimated mean by TF-IDF and one distant member per iteration is replaced with a closer one, say \begin{math} X_{m} \end{math} in \begin{math} i \end{math}th iteration, Then:

\begin{multline*}
X^{(i)}_{m} > X^{(i-1)}_{m} \Rightarrow \frac{X^{(i-1)}_{1} + X^{(i-1)}_{2} + X^{(i-1)}_{3} + ... + X^{(i-1)}_{m-1} + X^{(i)}_{m}}{m} > \\  \frac{X^{(i-1)}_{1} + X^{(i-1)}_{2} + X^{(i-1)}_{3} + ... + X^{(i-1)}_{m-1} + X^{(i-1)}_{m}}{m}
\end{multline*}
\begin{equation}
\end{equation}

and consequently:

\begin{multline*}
\mu^{(i)} > \mu^{(i-1)} > ... > \mu^{(1)} > \mu^{(0)} \Rightarrow ( \mu - \mu^{(i)}) < (\mu - \mu^{(i-1)}) < ... \\ < (\mu - \mu^{(1)}) < (\mu - \mu^{(0)})
\end{multline*}
\begin{equation}
\end{equation}

approaching the correct context's expected embedding \begin{math} \mu \end{math} in each iteration. In case of \begin{math} \mu < \mu^{(0)} \end{math}: 

\begin{multline*}
\mu^{(i)} < \mu^{(i-1)} < ... < \mu^{(1)} < \mu^{(0)} \\ \Rightarrow (\mu^{(i)} - \mu) < (\mu^{(i-1)} - \mu) < ... < (\mu^{(1)} - \mu) < (\mu^{(0)} - \mu)
\end{multline*}
\begin{equation}
\end{equation}

Finally, from (\ref{eq:6}):
\begin{equation}
\lim_{i \rightarrow \infty} \frac{X^{(i-1)}_{k}}{X^{(i)}_{k}} = 1 \Rightarrow \lim_{i \rightarrow \infty} \mid X^{(i)}_{k}-X^{(i-1)}_{k} \mid = 0 \Rightarrow \lim_{i \rightarrow \infty} \mid \mu^{(i)}-\mu^{(i-1)}\mid = 0
\end{equation}

and the condition of almost surely convergence is met after enough number of iterations:

\begin{equation}
    \left | \mu  - \mu^{(i)}\right | \leq \varepsilon 
\end{equation}

There are two ways that the proposed method can fail in improving the word rank. First, if the \begin{math} m \end{math}th and \begin{math} (m+1) \end{math}th words have exactly same score. In this case, the choice of words may change the expected embedding value and consequently lead to a totally different approximation of the document context. In order to solve this problem, the algorithm may simply check the score of \begin{math} m \end{math}th and the \begin{math} (m+1) \end{math}th words and in case of encountering the same scores, it can enter the \begin{math} (m+1) \end{math}th word into the set as well. Second, if before starting the refining process the conventional TF-IDF produces scores with very high error rate, STF-IDF fails to find the correct context. Nevertheless, our results show that such a high error rate is not a common case and a moderate performance of TF-IDF is enough for the current method to produce significantly good results.

\section{Results and discussion}

The algorithm was tested for 160 randomly chosen informal medical documents. For both conventional TF-IDF and STF-IDF, the scores were evaluated with human annotated labels. Since the precision of word importance can be subjective, here, we define and analyze the ranking error which measures the number of words were put between the first \begin{math} \left \lfloor \sqrt{n} \right \rfloor \end{math} most relevant terms while they have the least relevance based on human evaluation.

\begin{figure}[!ht]
  \centering
  \includegraphics[width=12cm]{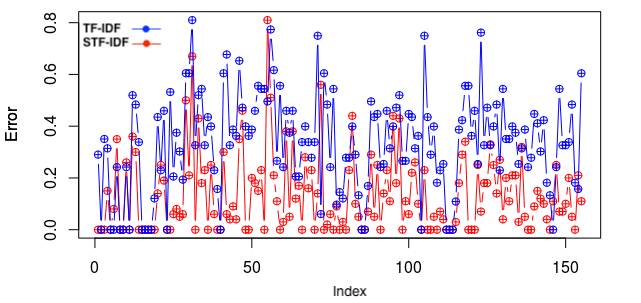}
  \caption{The error rate of STF-IDF in comparison with conventional TF-IDF for each document}
  \label{fig:lineError}
\end{figure}

As it is seen in Fig. \ref{fig:lineError}, by replacing better words in the ranking table, STF-IDF has less error rate in comparison with TF-IDF. Since STF-IDF is initially constructed upon the TF-IDF scores, in the rare cases (less than 5\% of times) when TF-IDF has abnormally big error rate, STF-IDF performs worse than TF-IDF.

\begin{figure}[!ht]
  \centering
  \includegraphics[width=10cm]{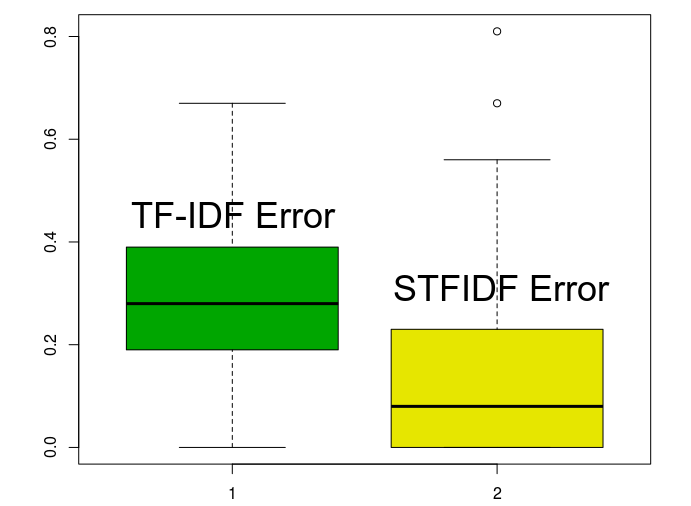}
  \caption{The boxplot of ranking error for STF-IDF and conventional TF-IDF}
  \label{fig:boxPlotError}
\end{figure}

We measured the error rate of STF-IDF against the original TF-IDF. As it is show in Fig. \ref{fig:boxPlotError}, our method improved the error rate by more than 50\%, decreasing the error rate of TF-IDF from 27.68\% to 13\%. Among the rankings generated for 160 documents, in 50\% of times the error rate of STF-IDF is significantly less ranging from 0\% to 13\% as opposed to high error of conventional TF-IDF ranging from 20\% to 30\% for 50\% of tested documents.
\newline
\newline
\newline
\newline
\newline
\newline

\section{Conclusion}

TF-IDF is an efficient and simple algorithm for matching words in a query to documents that are relevant to that query. From the data collected, TF-IDF returns documents that are highly relevant to a particular query. If a user were to input a query for a particular topic, TF-IDF can find documents that contain relevant information on the query. Furthermore, encoding TF-IDF is straightforward, making it ideal for forming the basis for more complicated algorithms and query retrieval systems \cite{Berger2000}.

Despite its strength, TF-IDF has its limitations. In terms of synonyms, notice that TF-IDF does not make the jump to the relationship between words. Going back to \cite{Berger1999}. In some experiments, TF-IDF could not equate the word .drug. with its plural .drugs., categorizing each instead as separate words and slightly decreasing the word.s wd value. For large document collections, this could present an escalating problem.

In this study we proposed a novel method based on semantically weighted TF-IDF scores for finding word relevance between a collection of documents. Textual data of nearly 4 million online medical communities were gathered and preprocessed. Afterwards, the corpus was fed to word2vec algorithm in order to generate word embedding. Initially, the words were ranked by conventional TF-IDF algorithm. Then these scores were repeatedly modified based on a semantic weight of each word proportional to the cosine distance of the world and the expected value of embedding of a set of most relevant words in each iteration. The algorithm stops when it reaches a predefined threshold which is a measure of dislocation of mean embedding distribution. Our results show a significant decrease in error rate when STF-IDF is utilized. The future works will be focused on the convergence proof of the algorithm as well as replacing automatic tests with human-involved evaluation.

\section{Conflict of interest}
The current method was proposed and tested by a group of data scientists from Itaú Unibanco. Any opinions, findings, and conclusions expressed in this manuscript are those of the authors and do not necessarily reflect the views, official policy or position of Itaú Unibanco.

%
%
%

\bibliographystyle{IEEEtran}

\bibliography{sample-bibliography}

\begin{thebibliography}{10}
\providecommand{\url}[1]{#1}
\csname url@samestyle\endcsname
\providecommand{\newblock}{\relax}
\providecommand{\bibinfo}[2]{#2}
\providecommand{\BIBentrySTDinterwordspacing}{\spaceskip=0pt\relax}
\providecommand{\BIBentryALTinterwordstretchfactor}{4}
\providecommand{\BIBentryALTinterwordspacing}{\spaceskip=\fontdimen2\font plus
\BIBentryALTinterwordstretchfactor\fontdimen3\font minus
  \fontdimen4\font\relax}
\providecommand{\BIBforeignlanguage}[2]{{%
\expandafter\ifx\csname l@#1\endcsname\relax
\typeout{** WARNING: IEEEtran.bst: No hyphenation pattern has been}%
\typeout{** loaded for the language `#1'. Using the pattern for}%
\typeout{** the default language instead.}%
\else
\language=\csname l@#1\endcsname
\fi
#2}}
\providecommand{\BIBdecl}{\relax}
\BIBdecl

\bibitem{zhang2006keyword}
K.~Zhang, H.~Xu, J.~Tang, and J.~Li, ``Keyword extraction using support vector
  machine,'' in \emph{International Conference on Web-Age Information
  Management}.\hskip 1em plus 0.5em minus 0.4em\relax Springer, 2006, pp.
  85--96.

\bibitem{uzun2005keyword}
Y.~Uzun, ``Keyword extraction using naive bayes,'' in \emph{Bilkent University,
  Department of Computer Science, Turkey www. cs. bilkent. edu. tr/\~{}
  guvenir/courses/CS550/Workshop/Yasin\_Uzun. pdf}, 2005.

\bibitem{Aizawa2003}
A.~Aizawa, ``An information-theoretic perspective of tf–idf measures,''
  \emph{Information Processing and Management}, vol.~39, no.~1, pp. 45--65,
  2003.

\bibitem{Salton1975}
G.~Salton, A.~Wong, and C.~S. Yang, ``A vector space model for automatic
  indexing,'' \emph{Commun. ACM}, vol.~18, p. 613–620, 1975.

\bibitem{almeida2019word}
F.~Almeida and G.~Xexéo, ``Word embeddings: A survey,'' \emph{arXiv}, vol.
  eprint 1901.09069, 2019.

\bibitem{Papineni2001}
K.~Papineni, ``Why inverse document frequency,'' \emph{Proceedings of the North
  American Association for Computational Linguistics}, pp. 25--32, 2001.

\bibitem{Singh2014}
J.~Singh and D.~S.K, ``Comparative analysis of idf methods to determine word
  relevance in web document,'' \emph{International Journal of Computer
  Science}, vol.~11, no.~1, pp. 1694--0784, 2014.

\bibitem{Ramos2003UsingTT}
J.~Ramos, ``Using tf-idf to determine word relevance in document queries,''
  2003.

\bibitem{Salton1988}
G.~Salton and C.~Buckley, ``Term-weighing approache sin automatic text
  retrieval,'' \emph{In Information Processing and Management}, vol.~24, pp.
  513--523, 1988.

\bibitem{Berger2000}
A.~Berger~et al, ``Bridging the lexical chasm: Statistical approaches to answer
  finding,'' \emph{Proc. Int. Conf. Research and Development in Information
  Retrieval}, pp. 192--199, 2000.

\bibitem{YJung2000}
Y.~Jung, H.~Park, and D.~Du, ``An effective term- weighting scheme for
  information retrieval,'' \emph{Technical Report TR00-008 Department of
  Computer Science and Engineering, University of Minnesota}, 2000.

\bibitem{ramos2003using}
J.~Ramos \emph{et~al.}, ``Using tf-idf to determine word relevance in document
  queries,'' in \emph{Proceedings of the first instructional conference on
  machine learning}, vol. 242, 2003, pp. 133--142.

\bibitem{li2007keyword}
J.~Li, K.~Zhang \emph{et~al.}, ``Keyword extraction based on tf/idf for chinese
  news document,'' \emph{Wuhan University Journal of Natural Sciences},
  vol.~12, no.~5, pp. 917--921, 2007.

\bibitem{wu2008interpreting}
H.~C. Wu, R.~W.~P. Luk, K.~F. Wong, and K.~L. Kwok, ``Interpreting tf-idf term
  weights as making relevance decisions,'' \emph{ACM Transactions on
  Information Systems (TOIS)}, vol.~26, no.~3, p.~13, 2008.

\bibitem{lee2008news}
S.~Lee and H.-j. Kim, ``News keyword extraction for topic tracking,'' in
  \emph{Networked Computing and Advanced Information Management, 2008. NCM'08.
  Fourth International Conference on}, vol.~2.\hskip 1em plus 0.5em minus
  0.4em\relax IEEE, 2008, pp. 554--559.

\bibitem{Ignacio2019}
I.~Arroyo-Fernández, C.-F. Méndez-Cruz, G.~Sierra, J.-M. Torres-Moreno, and
  G.~Sidorov, ``Unsupervised sentence representations as word information
  series: Revisiting tf–idf,'' \emph{Computer Speech and Language}, vol.~56,
  pp. 107--129, 2019.

\bibitem{Donghwa2019}
D.~Kim, D.~Seo, S.~Cho, and P.~Kang, ``Multi-co-training for document
  classification using various document representations: Tf–idf, lda, and
  doc2vec,'' \emph{Information Sciences}, vol. 477, pp. 15--29, 2019.

\bibitem{Iwendi2019}
C.~Iwendi, S.~Ponnan, R.~Munirathinam, K.~Srinivasan, and C.-Y. Chang, ``An
  efficient and unique tf/idf algorithmic model-based data analysis for
  handling applications with big data streaming,'' \emph{Electronics}, vol.~8,
  p. 1331, 2019.

\bibitem{morris2010comparison}
M.~R. Morris, J.~Teevan, and K.~Panovich, ``A comparison of information seeking
  using search engines and social networks.'' \emph{ICWSM}, vol.~10, pp.
  23--26, 2010.

\bibitem{wollmer2010bidirectional}
M.~W{\"o}llmer, F.~Eyben, A.~Graves, B.~Schuller, and G.~Rigoll,
  ``Bidirectional lstm networks for context-sensitive keyword detection in a
  cognitive virtual agent framework,'' \emph{Cognitive Computation}, vol.~2,
  no.~3, pp. 180--190, 2010.

\bibitem{mikolov2013efficient}
T.~Mikolov, K.~Chen, G.~Corrado, and J.~Dean, ``Efficient estimation of word
  representations in vector space,'' \emph{arXiv preprint arXiv:1301.3781},
  2013.

\bibitem{Berger1999}
A.~Berger and J.~Lafferty, ``Information retrieval as statistical
  translation,'' \emph{Proceedings of the 22nd ACM Conference on Research and
  Development in Information Retrieval}, pp. 222--229, 1999.

\end{thebibliography}


@INPROCEEDINGS{Gori2005, 
author={M. {Gori} and G. {Monfardini} and F. {Scarselli}}, 
booktitle={Proceedings. 2005 IEEE International Joint Conference on Neural Networks, 2005.}, 
title={A new model for learning in graph domains}, 
year={2005}, 
volume={2}, 
number={}, 
pages={729-734 vol. 2}, 
keywords={data structures;graph theory;neural nets;learning (artificial intelligence);graphical data structures;graph neural network;recursive neural networks;learning algorithm;Neural networks;Focusing;Application software;Machine learning;Recurrent neural networks;Encoding;Data structures;Machine learning algorithms;Tree graphs;Software engineering}, 
doi={10.1109/IJCNN.2005.1555942}, 
ISSN={2161-4393}, 
month={July},}

@article{Bengio2009learning,
  title={Learning deep architectures for AI},
  author={Bengio, Yoshua and others},
  journal={Foundations and trends{\textregistered} in Machine Learning},
  volume={2},
  number={1},
  pages={1--127},
  year={2009},
  publisher={Now Publishers, Inc.}
}

@ARTICLE{Hinton2012, 
author={G. {Hinton} and L. {Deng} and D. {Yu} and G. E. {Dahl} and A. {Mohamed} and N. {Jaitly} and A. {Senior} and V. {Vanhoucke} and P. {Nguyen} and T. N. {Sainath} and B. {Kingsbury}}, 
journal={IEEE Signal Processing Magazine}, 
title={Deep Neural Networks for Acoustic Modeling in Speech Recognition: The Shared Views of Four Research Groups}, 
year={2012}, 
volume={29}, 
number={6}, 
pages={82-97}, 
keywords={feedforward neural nets;Gaussian processes;hidden Markov models;speech recognition;deep neural networks;acoustic modeling;speech recognition;hidden Markov models;temporal variability;Gaussian mixture models;feed-forward neural network;posterior probabilities;HMM states;Automatic speech recognition;Speech recognition;Hidden Markov models;Training;Gaussian processes;Acoustics;Neural networks;Data models}, 
doi={10.1109/MSP.2012.2205597}, 
ISSN={1053-5888}, 
month={Nov},}

@inproceedings{Krizhevsky2012imagenet,
  title={Imagenet classification with deep convolutional neural networks},
  author={Krizhevsky, Alex and Sutskever, Ilya and Hinton, Geoffrey E},
  booktitle={Advances in neural information processing systems},
  pages={1097--1105},
  year={2012}
}

@article{Goodfellow2016NIPS2T,
  title={NIPS 2016 Tutorial: Generative Adversarial Networks},
  author={Ian J. Goodfellow},
  journal={CoRR},
  year={2016},
  volume={abs/1701.00160}
}

@book{Blanken2007multimedia,
  title={Multimedia retrieval},
  author={Blanken, Henk M and de Vries, Arjen P and Blok, Henk Ernst and Feng, Ling},
  year={2007},
  publisher={Springer}
}

@inproceedings{Karras2018progressive,
title={Progressive Growing of {GAN}s for Improved Quality, Stability, and Variation},
author={Tero Karras and Timo Aila and Samuli Laine and Jaakko Lehtinen},
booktitle={International Conference on Learning Representations},
year={2018},
url={https://openreview.net/forum?id=Hk99zCeAb},
}

@article{Berthelot2017BEGAN,
  title={BEGAN: Boundary Equilibrium Generative Adversarial Networks},
  author={David Berthelot and Tom Schumm and Luke Metz},
  journal={CoRR},
  year={2017},
  volume={abs/1703.10717}
}

@incollection{Jiajun2016NIPS,
title = {Learning a Probabilistic Latent Space of Object Shapes via 3D Generative-Adversarial Modeling},
author = {Wu, Jiajun and Zhang, Chengkai and Xue, Tianfan and Freeman, Bill and Tenenbaum, Josh},
booktitle = {Advances in Neural Information Processing Systems 29},
pages = {82--90},
year = {2016},
publisher = {Curran Associates, Inc.},
}

@inproceedings{Hamilton2017inductive,
  title={Inductive representation learning on large graphs},
  author={Hamilton, Will and Ying, Zhitao and Leskovec, Jure},
  booktitle={Advances in Neural Information Processing Systems},
  pages={1024--1034},
  year={2017}
}

@article{Kipf2016semi,
  title={Semi-supervised classification with graph convolutional networks},
  author={Kipf, Thomas N and Welling, Max},
  journal={arXiv preprint arXiv:1609.02907},
  year={2016}
} 

@article{Sanchez2018graph,
  title={Graph networks as learnable physics engines for inference and control},
  author={Sanchez-Gonzalez, Alvaro and Heess, Nicolas and Springenberg, Jost Tobias and Merel, Josh and Riedmiller, Martin and Hadsell, Raia and Battaglia, Peter},
  journal={arXiv preprint arXiv:1806.01242},
  year={2018}
}

@inproceedings{Battaglia2016interaction,
  title={Interaction networks for learning about objects, relations and physics},
  author={Battaglia, Peter and Pascanu, Razvan and Lai, Matthew and Rezende, Danilo Jimenez and others},
  booktitle={Advances in neural information processing systems},
  pages={4502--4510},
  year={2016}
}

@inproceedings{Fout2017protein,
  title={Protein interface prediction using graph convolutional networks},
  author={Fout, Alex and Byrd, Jonathon and Shariat, Basir and Ben-Hur, Asa},
  booktitle={Advances in Neural Information Processing Systems},
  pages={6530--6539},
  year={2017}
}

@article{Hamaguchi2017knowledge,
  title={Knowledge transfer for out-of-knowledge-base entities: A graph neural network approach},
  author={Hamaguchi, Takuo and Oiwa, Hidekazu and Shimbo, Masashi and Matsumoto, Yuji},
  journal={arXiv preprint arXiv:1706.05674},
  year={2017}
}

@inproceedings{Khalil2017learning,
  title={Learning combinatorial optimization algorithms over graphs},
  author={Khalil, Elias and Dai, Hanjun and Zhang, Yuyu and Dilkina, Bistra and Song, Le},
  booktitle={Advances in Neural Information Processing Systems},
  pages={6348--6358},
  year={2017}
}

@inproceedings{Bojchevski2018netgan,
  title={NetGAN: Generating Graphs via Random Walks},
  author={Bojchevski, Aleksandar and Shchur, Oleksandr and Z{\"u}gner, Daniel and G{\"u}nnemann, Stephan},
  booktitle={International Conference on Machine Learning},
  pages={609--618},
  year={2018}
}

@article{Zhou2018graph,
  title={Graph Neural Networks: A Review of Methods and Applications},
  author={Zhou, Jie and Cui, Ganqu and Zhang, Zhengyan and Yang, Cheng and Liu, Zhiyuan and Sun, Maosong},
  journal={arXiv preprint arXiv:1812.08434},
  year={2018}
}

@inproceedings{Hinneburg1998,
 author = {Hinneburg, Alexander and Keim, Daniel A.},
 title = {An Efficient Approach to Clustering in Large Multimedia Databases with Noise},
 booktitle = {Proceedings of the Fourth International Conference on Knowledge Discovery and Data Mining},
 series = {KDD'98},
 year = {1998},
 location = {New York, NY},
 pages = {58--65},
 numpages = {8},
 url = {http://dl.acm.org/citation.cfm?id=3000292.3000302},
 acmid = {3000302},
 publisher = {AAAI Press},
 keywords = {clustering algorithms, clustering in multimedia databases, clustering in the presence of noise, clustering of high-dimensional data, density-based clustering},
} 

@article{Zhou2009,
 author = {Zhou, Yang and Cheng, Hong and Yu, Jeffrey Xu},
 title = {Graph Clustering Based on Structural/Attribute Similarities},
 journal = {Proc. VLDB Endow.},
 issue_date = {August 2009},
 volume = {2},
 number = {1},
 month = aug,
 year = {2009},
 issn = {2150-8097},
 pages = {718--729},
 numpages = {12},
 url = {https://doi.org/10.14778/1687627.1687709},
 doi = {10.14778/1687627.1687709},
 acmid = {1687709},
 publisher = {VLDB Endowment},
} 

@article{Scarselli:2009:GNN:1657477.1657482,
 author = {Scarselli, Franco and Gori, Marco and Tsoi, Ah Chung and Hagenbuchner, Markus and Monfardini, Gabriele},
 title = {The Graph Neural Network Model},
 journal = {Trans. Neur. Netw.},
 issue_date = {January 2009},
 volume = {20},
 number = {1},
 month = jan,
 year = {2009},
 issn = {1045-9227},
 pages = {61--80},
 numpages = {20},
 url = {http://dx.doi.org/10.1109/TNN.2008.2005605},
 doi = {10.1109/TNN.2008.2005605},
 acmid = {1657482},
 publisher = {IEEE Press},
 address = {Piscataway, NJ, USA},
 keywords = {Graphical domains, graph neural networks (GNNs), graph processing, graphical domains, recursive neural networks},
} 

@article{DBLP:journals/corr/abs-1805-11724,
  author    = {Michael Kampffmeyer and
               Yinbo Chen and
               Xiaodan Liang and
               Hao Wang and
               Yujia Zhang and
               Eric P. Xing},
  title     = {Rethinking Knowledge Graph Propagation for Zero-Shot Learning},
  journal   = {CoRR},
  volume    = {abs/1805.11724},
  year      = {2018},
  url       = {http://arxiv.org/abs/1805.11724},
  archivePrefix = {arXiv},
  eprint    = {1805.11724},
  timestamp = {Mon, 13 Aug 2018 16:48:00 +0200},
  biburl    = {https://dblp.org/rec/bib/journals/corr/abs-1805-11724},
  bibsource = {dblp computer science bibliography, https://dblp.org}
}

@inproceedings{Zhang:2018:DCC:3178876.3186106,
 author = {Zhang, Yizhou and Xiong, Yun and Kong, Xiangnan and Li, Shanshan and Mi, Jinhong and Zhu, Yangyong},
 title = {Deep Collective Classification in Heterogeneous Information Networks},
 booktitle = {Proceedings of the 2018 World Wide Web Conference},
 series = {WWW '18},
 year = {2018},
 isbn = {978-1-4503-5639-8},
 location = {Lyon, France},
 pages = {399--408},
 numpages = {10},
 doi = {10.1145/3178876.3186106},
 acmid = {3186106},
 publisher = {International World Wide Web Conferences Steering Committee},
} 

@inproceedings{beck-etal-2018-graph,
    title = "Graph-to-Sequence Learning using Gated Graph Neural Networks",
    author = "Beck, Daniel  and
      Haffari, Gholamreza  and
      Cohn, Trevor",
    booktitle = "Proceedings of the 56th Annual Meeting of the Association for Computational Linguistics",
    month = jul,
    year = "2018",
    publisher = "Association for Computational Linguistics",
    pages = "273--283",
}

@inproceedings{Schlichtkrull2018ModelingRD,
  title={Modeling Relational Data with Graph Convolutional Networks},
  author={Michael Sejr Schlichtkrull and Thomas N. Kipf and Peter Bloem and Rianne van den Berg and Ivan Titov and Max Welling},
  booktitle={ESWC},
  year={2018}
}

@inproceedings{ae482107de73461787258f805cf8f4ed,
title = "Spectral networks and locally connected networks on graphs",
author = "Joan Bruna and Wojciech Zaremba and Arthur Szlam and Yann Lecun",
year = "2014",
language = "English (US)",
booktitle = "International Conference on Learning Representations (ICLR2014), CBLS, April 2014",
}

@incollection{NIPS2016_6212,
title = {Diffusion-Convolutional Neural Networks},
author = {Atwood, James and Towsley, Don},
booktitle = {Advances in Neural Information Processing Systems 29},
pages = {1993--2001},
year = {2016},
publisher = {Curran Associates, Inc.},
url = {http://papers.nips.cc/paper/6212-diffusion-convolutional-neural-networks.pdf}
}

@incollection{NIPS2017_6703,
title = {Inductive Representation Learning on Large Graphs},
author = {Hamilton, Will and Ying, Zhitao and Leskovec, Jure},
booktitle = {Advances in Neural Information Processing Systems 30},
pages = {1024--1034},
year = {2017},
publisher = {Curran Associates, Inc.},
url = {http://papers.nips.cc/paper/6703-inductive-representation-learning-on-large-graphs.pdf}
}

@article{Li2016GatedGS,
  title={Gated Graph Sequence Neural Networks},
  author={Yujia Li and Daniel Tarlow and Marc Brockschmidt and Richard S. Zemel},
  journal={CoRR},
  year={2016},
  volume={abs/1511.05493}
}

@inproceedings{tai-etal-2015-improved,
    title = "Improved Semantic Representations From Tree-Structured Long Short-Term Memory Networks",
    author = "Tai, Kai Sheng  and
      Socher, Richard  and
      Manning, Christopher D.",
    booktitle = "Proceedings of the 53rd Annual Meeting of the Association for Computational Linguistics and the 7th International Joint Conference on Natural Language Processing",
    month = jul,
    year = "2015",
    doi = "10.3115/v1/P15-1150",
    pages = "1556--1566",
}

@article{Bahdanau2015NeuralMT,
  title={Neural Machine Translation by Jointly Learning to Align and Translate},
  author={Dzmitry Bahdanau and Kyunghyun Cho and Yoshua Bengio},
  journal={CoRR},
  year={2015},
  volume={abs/1409.0473}
}

@inproceedings{gehring-etal-2017-convolutional,
    title = "A Convolutional Encoder Model for Neural Machine Translation",
    author = "Gehring, Jonas  and
      Auli, Michael  and
      Grangier, David  and
      Dauphin, Yann",
    booktitle = "Proceedings of the 55th Annual Meeting of the Association for Computational Linguistics",
    month = jul,
    year = "2017",
    doi = "10.18653/v1/P17-1012",
    pages = "123--135",
}

@incollection{NIPS2017_7181,
title = {Attention is All you Need},
author = {Vaswani, Ashish and Shazeer, Noam and Parmar, Niki and Uszkoreit, Jakob and Jones, Llion and Gomez, Aidan N and Kaiser, \L ukasz and Polosukhin, Illia},
booktitle = {Advances in Neural Information Processing Systems 30},
pages = {5998--6008},
year = {2017},
publisher = {Curran Associates, Inc.},
url = {http://papers.nips.cc/paper/7181-attention-is-all-you-need.pdf}
}

@inproceedings{cheng-etal-2016-long,
    title = "Long Short-Term Memory-Networks for Machine Reading",
    author = "Cheng, Jianpeng  and
      Dong, Li  and
      Lapata, Mirella",
    booktitle = "Proceedings of the 2016 Conference on Empirical Methods in Natural Language Processing",
    month = nov,
    year = "2016",
    doi = "10.18653/v1/D16-1053",
    pages = "551--561",
}

@article{velickovic2018graph,
  title="{Graph Attention Networks}",
  author={Veli{\v{c}}kovi{\'{c}}, Petar and Cucurull, Guillem and Casanova, Arantxa and Romero, Adriana and Li{\`{o}}, Pietro and Bengio, Yoshua},
  journal={International Conference on Learning Representations},
  year={2018},
  url={https://openreview.net/forum?id=rJXMpikCZ},
}

@INPROCEEDINGS{7780459, 
author={K. {He} and X. {Zhang} and S. {Ren} and J. {Sun}}, 
booktitle={2016 IEEE Conference on Computer Vision and Pattern Recognition (CVPR)}, 
title={Deep Residual Learning for Image Recognition}, 
year={2016}, 
pages={770-778}, 
doi={10.1109/CVPR.2016.90}, 
ISSN={1063-6919}, 
month={June},}

@InProceedings{pmlr-v80-xu18c,
  title = 	 {Representation Learning on Graphs with Jumping Knowledge Networks},
  author = 	 {Xu, Keyulu and Li, Chengtao and Tian, Yonglong and Sonobe, Tomohiro and Kawarabayashi, Ken-ichi and Jegelka, Stefanie},
  booktitle = 	 {Proceedings of the 35th International Conference on Machine Learning},
  pages = 	 {5453--5462},
  year = 	 {2018},
  month = 	 {10--15 Jul},
  publisher = 	 {PMLR},
  url = 	 {http://proceedings.mlr.press/v80/xu18c.html},
}

@inproceedings{chen2018fastgcn,
title={Fast{GCN}: Fast Learning with Graph Convolutional Networks via Importance Sampling},
author={Jie Chen and Tengfei Ma and Cao Xiao},
booktitle={International Conference on Learning Representations},
year={2018},
url={https://openreview.net/forum?id=rytstxWAW},
}

@article{DBLP:journals/corr/abs-1809-05343,
  author    = {Wen{-}bing Huang and
               Tong Zhang and
               Yu Rong and
               Junzhou Huang},
  title     = {Adaptive Sampling Towards Fast Graph Representation Learning},
  journal   = {CoRR},
  volume    = {abs/1809.05343},
  year      = {2018},
  url       = {http://arxiv.org/abs/1809.05343},
  archivePrefix = {arXiv},
  eprint    = {1809.05343},
  timestamp = {Fri, 05 Oct 2018 11:34:52 +0200},
  biburl    = {https://dblp.org/rec/bib/journals/corr/abs-1809-05343},
  bibsource = {dblp computer science bibliography, https://dblp.org}
}

@InProceedings{pmlr-v80-chen18p,
  title = 	 {Stochastic Training of Graph Convolutional Networks with Variance Reduction},
  author = 	 {Chen, Jianfei and Zhu, Jun and Song, Le},
  booktitle = 	 {Proceedings of the 35th International Conference on Machine Learning},
  pages = 	 {942--950},
  year = 	 {2018},
  month = 	 {10--15 Jul},
  publisher = 	 {PMLR},
  url = 	 {http://proceedings.mlr.press/v80/chen18p.html},
  
}

@article{DBLP:journals/corr/GilmerSRVD17,
  author    = {Justin Gilmer and
               Samuel S. Schoenholz and
               Patrick F. Riley and
               Oriol Vinyals and
               George E. Dahl},
  title     = {Neural Message Passing for Quantum Chemistry},
  year      = {2017},
  url       = {http://arxiv.org/abs/1704.01212},
  archivePrefix = {arXiv},
  eprint    = {1704.01212},
}

@article{DBLP:journals/corr/abs-1711-07971,
  author    = {Xiaolong Wang and
               Ross B. Girshick and
               Abhinav Gupta and
               Kaiming He},
  title     = {Non-local Neural Networks},
  journal   = {CoRR},
  volume    = {abs/1711.07971},
  year      = {2017},
  url       = {http://arxiv.org/abs/1711.07971},
  archivePrefix = {arXiv},
  eprint    = {1711.07971},
  timestamp = {Fri, 05 Apr 2019 07:29:46 +0200},
  biburl    = {https://dblp.org/rec/bib/journals/corr/abs-1711-07971},
  bibsource = {dblp computer science bibliography, https://dblp.org}
}

@article{DBLP:journals/corr/abs-1806-01261,
  author    = {Peter W. Battaglia and
               Jessica B. Hamrick and
               Victor Bapst and
               Alvaro Sanchez{-}Gonzalez and
               Vin{\'{\i}}cius Flores Zambaldi and
               Mateusz Malinowski and
               Andrea Tacchetti and
               David Raposo and
               Adam Santoro and
               Ryan Faulkner and
               {\c{C}}aglar G{\"{u}}l{\c{c}}ehre and
               Francis Song and
               Andrew J. Ballard and
               Justin Gilmer and
               George E. Dahl and
               Ashish Vaswani and
               Kelsey Allen and
               Charles Nash and
               Victoria Langston and
               Chris Dyer and
               Nicolas Heess and
               Daan Wierstra and
               Pushmeet Kohli and
               Matthew Botvinick and
               Oriol Vinyals and
               Yujia Li and
               Razvan Pascanu},
  title     = {Relational inductive biases, deep learning, and graph networks},
  journal   = {CoRR},
  year      = {2018},
  url       = {http://arxiv.org/abs/1806.01261},
  archivePrefix = {arXiv},
  eprint    = {1806.01261},
  
}

@article{DBLP:journals/corr/PerozziAS14,
  author    = {Bryan Perozzi and
               Rami Al{-}Rfou and
               Steven Skiena},
  title     = {DeepWalk: Online Learning of Social Representations},
  journal   = {CoRR},
  year      = {2014},
  url       = {http://arxiv.org/abs/1403.6652},
  archivePrefix = {arXiv},
  eprint    = {1403.6652},
  timestamp = {Mon, 13 Aug 2018 16:46:44 +0200},
  biburl    = {https://dblp.org/rec/bib/journals/corr/PerozziAS14},
  bibsource = {dblp computer science bibliography, https://dblp.org}
}

@article{DBLP:journals/corr/GroverL16,
  author    = {Aditya Grover and
               Jure Leskovec},
  title     = {node2vec: Scalable Feature Learning for Networks},
  journal   = {CoRR},
  year      = {2016},
  url       = {http://arxiv.org/abs/1607.00653},
  archivePrefix = {arXiv},
  eprint    = {1607.00653},
  timestamp = {Mon, 13 Aug 2018 16:48:14 +0200},
  biburl    = {https://dblp.org/rec/bib/journals/corr/GroverL16},
  bibsource = {dblp computer science bibliography, https://dblp.org}
}

@article{Kipf2016VariationalGA,
  title={Variational Graph Auto-Encoders},
  author={Thomas N. Kipf and Max Welling},
  journal={CoRR},
  year={2016},
  volume={abs/1611.07308}
}

@article{DBLP:journals/corr/abs-1711-08267,
  author    = {Hongwei Wang and
               Jia Wang and
               Jialin Wang and
               Miao Zhao and
               Weinan Zhang and
               Fuzheng Zhang and
               Xing Xie and
               Minyi Guo},
  title     = {GraphGAN: Graph Representation Learning with Generative Adversarial
               Nets},
  journal   = {CoRR},
  year      = {2017},
  url       = {http://arxiv.org/abs/1711.08267},
  archivePrefix = {arXiv},
  eprint    = {1711.08267},
  timestamp = {Sun, 21 Apr 2019 10:04:41 +0200},
  biburl    = {https://dblp.org/rec/bib/journals/corr/abs-1711-08267},
  bibsource = {dblp computer science bibliography, https://dblp.org}
}

@article{DBLP:journals/corr/LiuCCOYS17,
  author    = {Weiyi Liu and
               Pin{-}Yu Chen and
               Hal Cooper and
               Min Hwan Oh and
               Sailung Yeung and
               Toyotaro Suzumura},
  title     = {Can {GAN} Learn Topological Features of a Graph?},
  journal   = {CoRR},
  year      = {2017},
  url       = {http://arxiv.org/abs/1707.06197},
  archivePrefix = {arXiv},
  eprint    = {1707.06197},
  timestamp = {Mon, 13 Aug 2018 16:47:20 +0200},
  biburl    = {https://dblp.org/rec/bib/journals/corr/LiuCCOYS17},
  bibsource = {dblp computer science bibliography, https://dblp.org}
}

@article{Tavakoli2017,
author = {Tavakoli, Sahar and Hajibagheri, Alireza and Sukthankar, Gita},
year = {2017},
month = {07},
title = {Learning Social Graph Topologies using Generative Adversarial Neural Networks},
doi = {10.13140/RG.2.2.16772.94082}
}

@article{DBLP:journals/corr/YuZWY16,
  author    = {Lantao Yu and
               Weinan Zhang and
               Jun Wang and
               Yong Yu},
  title     = {SeqGAN: Sequence Generative Adversarial Nets with Policy Gradient},
  journal   = {CoRR},
  year      = {2016},
  url       = {http://arxiv.org/abs/1609.05473},
  archivePrefix = {arXiv},
  eprint    = {1609.05473},
  timestamp = {Sun, 21 Apr 2019 10:04:41 +0200},
  biburl    = {https://dblp.org/rec/bib/journals/corr/YuZWY16},
  bibsource = {dblp computer science bibliography, https://dblp.org}
}

@article{articlekusner,
author = {J. Kusner, Matt and Miguel Hernández-Lobato, José},
year = {2016},
month = {11},
pages = {},
archivePrefix = {arXiv},
eprint    = {1611.04051},
title = {GANS for Sequences of Discrete Elements with the Gumbel-softmax Distribution}
}

@article{DBLP:journals/corr/LiMSRJ17,
  author    = {Jiwei Li and
               Will Monroe and
               Tianlin Shi and
               Alan Ritter and
               Dan Jurafsky},
  title     = {Adversarial Learning for Neural Dialogue Generation},
  journal   = {CoRR},
  year      = {2017},
  url       = {http://arxiv.org/abs/1701.06547},
  archivePrefix = {arXiv},
  eprint    = {1701.06547},
  timestamp = {Mon, 13 Aug 2018 16:46:30 +0200},
  biburl    = {https://dblp.org/rec/bib/journals/corr/LiMSRJ17},
  bibsource = {dblp computer science bibliography, https://dblp.org}
}

@article{DBLP:journals/corr/LiangHZGX17,
  author    = {Xiaodan Liang and
               Zhiting Hu and
               Hao Zhang and
               Chuang Gan and
               Eric P. Xing},
  title     = {Recurrent Topic-Transition {GAN} for Visual Paragraph Generation},
  journal   = {CoRR},
  year      = {2017},
  url       = {http://arxiv.org/abs/1703.07022},
  archivePrefix = {arXiv},
  eprint    = {1703.07022},
  timestamp = {Wed, 12 Dec 2018 16:25:50 +0100},
  biburl    = {https://dblp.org/rec/bib/journals/corr/LiangHZGX17},
  bibsource = {dblp computer science bibliography, https://dblp.org}
}

@article{DBLP:journals/corr/CheLZHLSB17,
  author    = {Tong Che and
               Yanran Li and
               Ruixiang Zhang and
               R. Devon Hjelm and
               Wenjie Li and
               Yangqiu Song and
               Yoshua Bengio},
  title     = {Maximum-Likelihood Augmented Discrete Generative Adversarial Networks},
  journal   = {CoRR},
  year      = {2017},
  url       = {http://arxiv.org/abs/1702.07983},
  archivePrefix = {arXiv},
  eprint    = {1702.07983},
  timestamp = {Mon, 13 Aug 2018 16:48:45 +0200},
  biburl    = {https://dblp.org/rec/bib/journals/corr/CheLZHLSB17},
  bibsource = {dblp computer science bibliography, https://dblp.org}
}

@inproceedings{devon2018boundary,
title={Boundary Seeking {GAN}s},
author={R Devon Hjelm and Athul Paul Jacob and Adam Trischler and Gerry Che and Kyunghyun Cho and Yoshua Bengio},
booktitle={International Conference on Learning Representations},
year={2018},
url={https://openreview.net/forum?id=rkTS8lZAb},
}

@article{mccallum2000automating,
  title={Automating the construction of internet portals with machine learning},
  author={McCallum, Andrew Kachites and Nigam, Kamal and Rennie, Jason and Seymore, Kristie},
  journal={Information Retrieval},
  volume={3},
  number={2},
  pages={127--163},
  year={2000},
  publisher={Springer}
}

[1]  Y. Bengioet al., "Learning deep architectures for ai,"Foundations andtrends® in Machine Learning, vol. 2, no. 1, pp. 1–127, 2009.

[2]  G.  Hinton,  L.  Deng,  D.  Yu,  G.  E.  Dahl,  A.  Mohamed,  N.  Jaitly,A. Senior, V. Vanhoucke, P. Nguyen, T. N. Sainath, and B. Kingsbury,"Deep  neural  networks  for  acoustic  modeling  in  speech  recognition:The  shared  views  of  four  research  groups,"IEEE  Signal  ProcessingMagazine, vol. 29, no. 6, pp. 82–97, Nov 2012.

[3]  A. Krizhevsky, I. Sutskever, and G. E. Hinton, "Imagenet classificationwith deep convolutional neural networks," inAdvances in neural infor-mation processing systems, 2012, pp. 1097–1105.

[4]  J.  Zhou,  G.  Cui,  Z.  Zhang,  C.  Yang,  Z.  Liu,  and  M.  Sun,  "Graphneural networks: A review of methods and applications,"arXiv preprintarXiv:1812.08434, 2018.

[5]  W.   Hamilton,   Z.   Ying,   and   J.   Leskovec,   "Inductive   representationlearning on large graphs," inAdvances in Neural Information ProcessingSystems, 2017, pp. 1024–1034.

[6]  T. N. Kipf and M. Welling, "Semi-supervised classification with graphconvolutional networks,"arXiv preprint arXiv:1609.02907, 2016.

[7]  A. Sanchez-Gonzalez, N. Heess, J. T. Springenberg, J. Merel, M. Ried-miller,   R.   Hadsell,   and   P.   Battaglia,   "Graph   networks   as   learn-able   physics   engines   for   inference   and   control,"arXiv   preprintarXiv:1806.01242, 2018.

[8]  P.  Battaglia,  R.  Pascanu,  M.  Lai,  D.  J.  Rezendeet  al.,  "Interactionnetworks for learning about objects, relations and physics," inAdvancesin neural information processing systems, 2016, pp. 4502–4510.

[9]  A.   Fout,   J.   Byrd,   B.   Shariat,   and   A.   Ben-Hur,   "Protein   interfaceprediction using graph convolutional networks," inAdvances in NeuralInformation Processing Systems, 2017, pp. 6530–6539.

[10]  T.  Hamaguchi,  H.  Oiwa,  M.  Shimbo,  and  Y.  Matsumoto,  "Knowledgetransfer  for  out-of-knowledge-base  entities:  A  graph  neural  networkapproach,"arXiv preprint arXiv:1706.05674, 2017.

[11]  E.  Khalil,  H.  Dai,  Y.  Zhang,  B.  Dilkina,  and  L.  Song,  "Learningcombinatorial  optimization  algorithms  over  graphs,"  inAdvances  inNeural Information Processing Systems, 2017, pp. 6348–6358.

[12]  M.  Gori,  G.  Monfardini,  and  F.  Scarselli,  "A  new  model  for  learningin  graph  domains,"  inProceedings.  2005  IEEE  International  JointConference on Neural Networks, 2005., vol. 2, July 2005, pp. 729–734vol. 2.

[13]  F.Scarselli,M.Gori,A.C.Tsoi,M.Hagenbuchner,andG.   Monfardini,   "The   graph   neural   network   model,"Trans.   Neur.Netw.,   vol.   20,   no.   1,   pp.   61–80,   Jan.   2009.   [Online].   Available:http://dx.doi.org/10.1109/TNN.2008.2005605

[14]  I. J. Goodfellow, "Nips 2016 tutorial: Generative adversarial networks,"CoRR, vol. abs/1701.00160, 2016.

[15]  H.  M.  Blanken,  A.  P.  de  Vries,  H.  E.  Blok,  and  L.  Feng,Multimediaretrieval.    Springer, 2007.

[16]  T.  Karras,  T.  Aila,  S.  Laine,  and  J.  Lehtinen,  "Progressive  growing  ofGANs  for  improved  quality,  stability,  and  variation,"  inInternationalConference  on  Learning  Representations,  2018.  [Online].  Available:https://openreview.net/forum?id=Hk99zCeAb

[17]  D. Berthelot, T. Schumm, and L. Metz, "Began: Boundary equilibriumgenerative adversarial networks,"CoRR, vol. abs/1703.10717, 2017.

[18]  J. Wu, C. Zhang, T. Xue, B. Freeman, and J. Tenenbaum, "Learning aprobabilistic latent space of object shapes via 3d generative-adversarialmodeling,"  inAdvances  in  Neural  Information  Processing  Systems  29.Curran Associates, Inc., 2016, pp. 82–90.

[19]  A.  Bojchevski,  O.  Shchur,  D.  Zügner,  and  S.  Günnemann,  "Netgan:Generating  graphs  via  random  walks,"  inInternational  Conference  onMachine Learning, 2018, pp. 609–618.

[20]  A.  Hinneburg  and  D.  A.  Keim,  "An  efficient  approach  to  clustering  inlarge  multimedia  databases  with  noise,"  inProceedings  of  the  FourthInternational  Conference  on  Knowledge  Discovery  and  Data  Mining,ser.  KDD'98.AAAI  Press,  1998,  pp.  58–65.  [Online].  Available:http://dl.acm.org/citation.cfm?id=3000292.3000302

[21]  Y.Zhou,H.Cheng,andJ.X.Yu,"Graphclusteringbased    on    structural/attribute    similarities,"Proc.    VLDB    Endow.,vol.    2,    no.    1,    pp.    718–729,    Aug.    2009.    [Online].    Available:https://doi.org/10.14778/1687627.1687709

[22]  A. K. McCallum, K. Nigam, J. Rennie, and K. Seymore, "Automatingthe construction of internet portals with machine learning,"InformationRetrieval, vol. 3, no. 2, pp. 127–163, 2000.

\vspace{12pt}

\end{document}